# Cytoplasmic flow induced by a rotating wire in living cells: Magnetic rotational spectroscopy and finite element simulations

**Charles Paul Moore[1], Foad Ghasemi[2] and Jean-François Berret[1*]**

[1]*Université Paris Cité, CNRS, Matière et Systèmes Complexes, 75013 Paris, France*
[2]*Institut Pasteur, INSERM UMR1225, Pathogenesis of Vascular Infections, 75015 Paris, France*

**Abstract**: Recent studies have highlighted intracellular viscosity as a key biomechanical property with potential as a biomarker for cancer cell metastasis. In the context of cellular mechanobiology, magnetic rotational spectroscopy (MRS), which employs rotating magnetic wires of length $L$ = 2-8 μm to probe cytoplasmic rheology, has emerged as an effective method for quantifying intracellular viscoelasticity. This study examines microrheology data from three breast epithelial cell lines, MCF-10A, MCF-7, and MDA-MB-231, along with new data from HeLa cervical cancer cells. Here, MRS is combined with finite element simulations to characterize the flow field induced by wire rotation in the cytoplasm. COMSOL simulations performed at low Reynolds numbers show that the flow velocity is localized around the wire, and display characteristic dumbbell-shaped profiles. For wires representative of MRS experiments in cells, the product of shear rate and cytoplasmic relaxation time ($\dot{\gamma}\tau$ with $\tau \sim 1$ s) remains below unity, indicating that the flow occurs within the linear regime. This outcome confirms that MRS can reliably measure the zero-shear viscosity of the intracellular medium in living cells. This study also demonstrates that integrating MRS intracellular measurements with COMSOL simulations significantly improves the reliability of *in vitro* assessments of cytoplasmic mechanical properties.

# I – Introduction

One objective of studying cell biomechanics is to identify physical biomarkers that correlate with cancer cell metastatic potential, thus contributing to the development of predictive diagnostic tools [1-4]. In the context of cell and tissue mechanobiology, three characteristic length scales are generally distinguished: the intracellular scale (< 10 µm), the whole-cell scale (10–30 µm), and the tissue scale (> 30 µm). A recent meta-analysis covering twenty years of research on cancer cell biomechanics [4] reveals that over 75% of studies have focused on the whole-cell scale, primarily to assess cell stiffness $K$ or apparent Young modulus $E_{App}$ [5]. Among the various techniques developed [6-8], atomic force microscopy (AFM) has emerged as the dominant approach, accounting for more than 90% of studies in the field [9]. As a result, a substantial body of data has accumulated on the mechanical properties of normal and cancerous cells across multiple cancer types, including breast [10-16], bladder [17-20], prostate [21-24], pancreatic [25, 26] and ovarian cancers [27]. These data consistently indicate that cancer cells are softer than their normal counterparts, showing a decrease in Young modulus ranging from 50% to 80% depending on the cancer type [9]. However, a major limitation of mechanobiology measurements lies in the substantial variability found across studies. A striking example is provided by AFM measurements on MCF-10A breast epithelial cells, where reported values of the apparent Young modulus vary by over two orders of magnitude, ranging from $E_{App}$ = 0.34 kPa [28] to 37.5 kPa [29]. This large dispersion is not limited to whole-cell measurements; it is also observed at the intracellular level and within cancer cells themselves [24, 30-33]. These issues highlight a central challenge in cancer cell biomechanics: the difficulty of obtaining absolute and reproducible values for rheological parameters. This stands in contrast to conventional rheometry, which has provided reliable rheological data for a wide variety of fluids and materials, including polymers, colloids and surfactants [34]. This article aims to address this challenge by combining magnetic rotational spectroscopy (MRS), an active microrheology technique [35], with COMSOL-based finite element flow simulations [36].

Although the viscoelastic nature of cells has long been acknowledged [37], most studies in cell mechanics have focused on elasticity, as previously mentioned, while viscosity has received much less attention. According to a recent review [9], only about 25% of investigations comparing cancerous and healthy cells have explored intracellular and whole-cell viscosity properties. The most commonly used techniques for assessing intracellular viscosity include particle-tracking microrheology, micropipette aspiration, optical tweezers and more recently magnetic rotational spectroscopy (MRS) [9]. In MRS, micrometer-sized magnetic wires act as actuators, driven by an external rotating magnetic field. By analyzing the rotational behavior of the wire under optical microscopy as a function of both angular frequency $\omega$ and wire length $L$, both viscosity and elastic modulus can be quantified [38, 39]. In this context, we investigated breast epithelial cell lines, such as MCF-10A, MCF-7 and MDA-MB-231, with the latter two representing low and high metastatic potential, respectively [39]. The most significant finding of this study was the correlation between cytoplasmic viscosity and metastatic activity, with a six-fold decrease observed between MCF-7 and MDA-MB-231 cells. This result suggested that cytoplasmic viscosity could serve as a potential mechanical marker for malignant cancer cells.

The Maxwell model, whether in its simplest form with a single relaxational mode or in its generalized form with multiple modes, is well suited to describe viscoelastic liquids. A Maxwell fluid is defined by two key parameters: the zero-shear viscosity $\eta_0$ and the instantaneous shear modulus $G_0$ [40]. Their ratio, $\tau = \eta_0/G_0$ defines the characteristic relaxation time of the medium, which is assessed from step strain experiments [34, 41]. This timescale reflects specific physical processes that, at the microscopic level, are responsible for relaxing the stress induced by the applied deformation, e.g. reptation in polymer melts or cage rearrangement in dense colloidal dispersions [42, 43]. In the linear shear regime defined by $\dot{\gamma}\tau \ll 1$, the Maxwell model predicts a steady-state shear stress $\sigma(\dot{\gamma}) = \eta_0\dot{\gamma}$, indicating direct proportionality. Conversely, in the non-linear flow regime where $\dot{\gamma}\tau \gg 1$, the apparent viscosity $\sigma(\dot{\gamma})/\dot{\gamma}$ becomes shear-rate-dependent, exhibiting either an increase (shear-thickening) or a decrease (shear-thinning) depending on the induced microstructure [34]. The purpose of this article is to revisit MCF-10A, MCF-7 and MDA-MB-231 intracellular viscosity data from Ref. [39], together with new, previously unpublished data on HeLa cervical cancer cells. In addition, it aims to characterize the deformations, shear rates, and sheared volumes induced by the rotation of a magnetic wire embedded in the cytoplasm, quantities that are normally accessible through conventional rheometry and are central to rheological analysis. To this end, we developed COMSOL finite element simulations to model the flow around the rotating wire. Our objective is also to determine whether the measured viscosity corresponds to a linear or nonlinear flow regime and, ultimately to assess whether MRS can reliably measure the intrinsic rheological properties of the cytoplasm.

## II - Materials and Methods

### II.1 – Magnetic wire synthesis and characterization

Magnetic wires used in this study are synthesized through electrostatic co-assembly of 13.2 nm poly(acrylic acid)-coated iron oxide nanoparticles [44] and cationic

poly(diallyldimethylammonium chloride) polymers (Sigma-Aldrich, Saint-Quentin-Fallavier, France). The assembly is triggered by a desalting transition and driven by a 0.3 T magnetic field, producing superparamagnetic wires with lengths ranging from 1 to 100 μm and diameters between 0.2 and 2 μm. For cell microrheology, the wires are sonicated to reduce their size to 1–10 μm [38]. Their morphology and composition are characterized by scanning electron microscopy and energy-dispersive X-ray spectroscopy [39], confirming the presence of iron-rich and rigid structures required for MRS.

## II.2 – Optical microscopy and environment

Wire rotation was monitored using an Olympus IX73 inverted microscope equipped with a 60× objective (NA 0.70), allowing bright-field and phase-contrast imaging. Images were captured with an EXi Blue CCD camera and processed using ImageJ. A custom-built magnetic device generated a rotating field (12 mT) via two orthogonal coil pairs driven with a 90° phase shift. The system allowed frequency sweeps from $2 \times 10^{-3}$ to 100 rad s$^{-1}$. Experiments were performed at 37 °C using an air stream for thermal stabilization. This setup enabled precise tracking of wire dynamics under controlled magnetic actuation (Supporting Information S1).

## II.3 – Cell cultures

MCF-10A cells (ATCC CRL-10317) were grown in T25 flasks as adherent monolayers using DMEM/F12-GlutaMAX medium (Gibco). The basal medium was supplemented with 5% horse serum, 1% penicillin–streptomycin, and 1% MEGM supplement (Lonza), the latter containing bovine pituitary extract (0.4%), recombinant human insulin-like growth factor-I (0.01 μg mL$^{-1}$), hydrocortisone (0.5 μg mL$^{-1}$), and human epidermal growth factor (3 ng mL$^{-1}$). MCF-7 cells (ATCC

HTB-22) were maintained in DMEM supplemented with 10% fetal bovine serum (FBS) and 1% penicillin–streptomycin (PAA Laboratories GmbH). MDA-MB-231 cells (ATCC HTB-26) were cultured under similar conditions, except that DMEM with high glucose (4.5 g $L^{-1}$) was used. HeLa cells (ATCC CCL-2) were grown in DMEM with high glucose (4.5 g $L^{-1}$) containing stable glutamine (PAA Laboratories GmbH, Austria), supplemented with 10% FBS and 1% penicillin–streptomycin. All cultures were maintained at 37 °C in a humidified atmosphere of 5% $CO_2$ and subcultured twice per week. Before passage, cells were washed with PBS and detached using trypsin–EDTA (PAA Laboratories GmbH). After centrifugation at 1200 g for 5 min, pellets were resuspended in assay medium, and cell numbers were determined using a Malassez chamber. Experiments with magnetic wires were performed at 60–80% confluence.

**II.4 – Magnetic rotational spectroscopy measuring protocol**

This protocol outlines the preparation, incubation, and measurement steps required to quantify intracellular viscosity and elasticity using magnetic wires actuated by a rotating magnetic field. Magnetic wires are first prepared in aqueous dispersion and imaged under bright-field microscopy to estimate their concentration and length distribution. Typical values range from $2\times10^4$ to $2\times10^5$ wires per µL. Dispersions are sterilized by autoclaving and stored in the refrigerator. Cells are seeded on ethanol-cleaned 30 mm glass coverslips placed in 6-well plates equipped with silicone spacers. Depending on the confluence needed for the experiments, seeding densities range from $1\times10^5$ to $6\times10^5$ cells per well. A 24-hour incubation period is typically allowed for the wires to be spontaneously uptaken by the cells [45]. The suspension is diluted to achieve approximately one to two wires per cell to optimize measurement throughput while ensuring uniform distribution. On the day of the experiment, cells are mounted in a thermostated chamber (PeCon,

Germany) filled with HEPES-buffered medium to maintain pH stability outside the incubator. After assembly, the chamber is placed on a microscope stage equipped with a custom-built magnetic device. The temperature is maintained at 36–37 °C throughout. Viscosity measurements are performed by reducing the excitation frequency in steps from 1 to $10^{-3}$ rad $s^{-1}$, until the synchronous regime is reached. Time-lapse movies are recorded at fixed angular frequencies $\omega$ = $1.7 \times 10^{k}$, $2.9 \times 10^{k}$ $5.3 \times 10^{k}$ and $9.4 \times 10^{k}$ rad $s^{-1}$, with k = -3 to 1. Wire angle, center-of-mass and length as a function of the time are retrieved using a home-made plugin in ImageJ [46]. From these recordings, the average rotational velocities and transition frequencies are extracted, providing access to the cytoplasm viscosity. Elasticity measurements are performed first at high actuation frequencies (100 rad $s^{-1}$) to assess the amplitude of wire oscillations. Fast acquisition settings (exposure time 1–10 ms) and bright-field imaging are used to preserve contrast and resolve rapid movements [39]. The wires are calibrated in water–glycerol solutions, which provides a viscosity measurement accuracy of about 14% [38, 47].

### IV.5 – Computational fluid mechanics: COMSOL Multiphysics finite element simulations

The flow field surrounding the magnetic wire was simulated using the COMSOL Multiphysics computational fluid dynamics environment (Version 6.2, license number 8079561) [36]. Specifically, the simulation involved the steady state solution in the laminar flow module. For wires of lengths 3.5 µm, the velocity fields were first calculated using simulation cubic domains with side lengths ranging from 10 to 500 µm. The data showed good overlap of the flow velocity when domains extended to at least 20 times the wire length. This value has been maintained throughout this study. The fluid material was set to non-solid, with density and dynamic viscosity specified. A pressure point constraint was placed in the corner of the volume to zero the field. The exterior walls of the surrounding cube are set to no slip condition with 0 velocity, i.e. assuming no external flow. The cylinder walls are set to no slip condition, with a surface velocity set to $v_x = -y\omega$ and $v_y = x\omega$, $v_z = 0$, where the positions $x$ and $y$ axes are centred in the centre of the wire. The steady state flow field was calculated using default COMSOL physics settings, corresponding to a "frozen" flow field. The shear field was determined using the flow field data. Meshing is based on the built-in COMSOL physics mesh algorithm, set to extra fine, with a further refinement near the wire. Meshing was selected to provide adequately granular flow field results in the vicinity of the wire.

# III – Results and discussion

## III.1 – Overview of magnetic rotational spectroscopy results in living cells

### *III.1.1 – Intracellular Viscosity and Elasticity*

This section provides an overview of the fundamental principles of the MRS technique [48] and summarizes key findings obtained from four eukaryotic cell lines [38, 39]. These insights will serve as the foundation for the finite element simulations presented in the subsequent sections. MRS was first developed to investigate the behavior of anisotropic, micron-sized magnetic objects embedded in a purely viscous liquid and exposed to a rotating magnetic field [49, 50]. The model was later extended by us to viscoelastic liquids and soft solids [47, 51]. Here, we outline the key aspects of MRS as applied to living cells, focusing on their rheological properties, specifically the shear viscosity $\eta_0$ and elastic modulus $G_0$. In Newtonian and Maxwell-type viscoelastic fluids, a magnetic wire (length $L$, diameter $D$) subjected to a rotating field undergoes a transition from a synchronous (S) to an asynchronous (AS) regime, the latter characterized by oscillations [52, 53]. The critical frequency governing this transition is given by [38, 48]:

$$\omega_C = \frac{3}{8\mu_0}\frac{\Delta\chi}{\eta_0}\frac{B^2}{L^{*2}} \quad with \ L^* = L\Big/\Big[D\sqrt{g(L/D)}\Big] \tag{1}$$

where $\Delta\chi$ denotes the anisotropy of susceptibility, $\mu_0$ is the vacuum permittivity, $B$ the magnetic field, $L^*$ the reduced length and $g(p) = ln(p) - 0.662 + 0.917p - 0.050p^2$ [47]. The anisotropy of magnetic susceptibility $\Delta\chi$ = 1.37 ± 0.09 was determined from $\omega_C(L^*)$-measurements using a water–glycerol sample of known viscosity (0.164 Pa s at 34 °C). To improve accuracy, intracellular viscosity is determined by analyzing a large dataset of $\omega_C(L)$ for each cell line. Approximately one hundred wires located in different cells are tracked while the rotation frequency varies from $10^{-3}$ to $10^2$ rad s$^{-1}$. The inherent wire size distribution ($L = 2 - 8$ μm) also enables us to determine the dependence $\omega_C(L^*)$ and verify it against the prediction of Eq. 1. In viscosity measurements, accessing low frequencies down to $10^{-3}$ rad s$^{-1}$ is essential to reveal the viscous effects of intracellular flow. To position MRS relative to widely used biomechanical methods, Fig. 1 presents an overview of the accessible frequency and time ranges. These techniques differ in how they probe intracellular properties—such as particle tracking, optical tweezers, and magnetic rotational spectroscopy—or whole-cell mechanics, as in atomic force microscopy, micropipette aspiration, magnetic tweezers, and microplate stretching. In particular, MRS extends to much lower frequencies than other techniques.

MRS is particularly effective in distinguishing a viscoelastic liquid from a soft solid, a class of materials characterized by a nonzero yield stress [34]. In the latter case, no S-AS transition occurs, and the wire remains in an oscillatory mode regardless of the applied frequency [51]. As wires in cells consistently exhibit a S-AS transitions, it was concluded [38] that the intracellular medium of NIH/3T3 fibroblasts and Hela cells exhibits viscoelastic liquid behavior and can be described as a generalized Maxwell fluid [34, 54]. Furthermore, the technique also allows differentiation between Newtonian and Maxwell fluid models by analyzing the amplitude of oscillation $\theta_B(\omega)$ in the asynchronous regime. In a purely viscous liquid, for $\omega > \omega_C$, $\theta_B(\omega)$ decreases as $1/\omega$ and tends hence toward zero at high frequencies. In contrast, for a viscoelastic fluid, $\theta_B(\omega)$ remains finite in the high-frequency limit and scales inversely with the elastic modulus $G_0$ [47]:

$$\lim_{\omega \to \infty} \theta_B(\omega) = \theta_0 = \frac{3}{4\mu_0} \frac{\Delta\chi}{G_0} \frac{B^2}{L^{*2}} \qquad (2)$$

In elasticity experiments, the applied frequencies are in the 100 rad s$^{-1}$ range, with the constraint that $\omega/\omega_C > 1000$, as discussed in Ref. [39]. From the original wire selection, this requirement ultimately limits the number of effective wires available for $G_0$-measurement to 40–50.

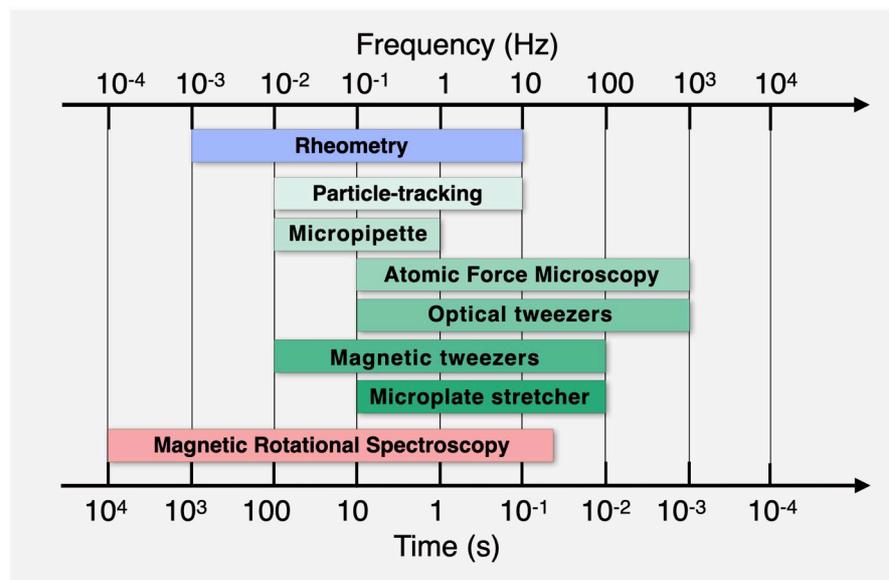

**Figure 1:** Comparison diagram of the frequency (upper scale in Hertz) and time (lower scale in second) ranges accessible by the most commonly used cellular microrheology techniques: particle tracking, micropipette aspiration, atomic force microcopy, optical and magnetic tweezers, microplate stretcher and magnetic rotational spectroscopy [9]. These techniques can probe either the cell interior or the whole cell, measuring viscosity and/or elastic modulus. For comparison, the performance of conventional rheometry, which uses rheometers and sample volumes on the order of milliliters, is also indicated.

*III.1.2 – Overview of findings on Adherent Eukaryotic Cell Lines using MRS*

Among the studied cell lines, we selected the following as representative examples of this class of biological materials: the breast epithelial cell lines MCF-10A, MCF-7, and MDA-MB-231, the latter two being tumorigenic with varying degrees of invasiveness, and HeLa cervical cancer cells. An example of a microscopy image of a HeLa cell monolayer incubated with magnetic wires, acquired at 60× magnification, is shown in Supporting Information S2. To establish representative simulation parameters, we selected those corresponding to an average figure on the basis of a statistical analysis of the wires internalized in the four cell lines. This analysis revealed lengths ranging from 2 to 8 μm, with a peak distribution around 3–4 μm and an aspect ratio of approximately 5 (Supporting Information S3). Based on these findings, initial simulations were performed for magnetic wires with a length of 3.5 μm and a diameter of 0.70 μm. In the simulations, we focus on the synchronous regime, where the wire rotates in phase with the field at an angular frequency $\Omega_S(\omega) = \omega$, which is bounded by $\omega_C$. Beyond this threshold, the wire average rotational velocity denoted as $\Omega_{AS}(\omega)$ decreases with increasing frequency, satisfying also the condition $\Omega_{AS}(\omega) < \omega_C$ [38, 47]. For this reason, the experimental $\omega_C$ values measured in cells will be used as the reference rotation frequency in our simulations. Table I presents the critical frequency $\omega_C$, viscosity $\eta_0$, elastic modulus $G_0$, and relaxation time $\tau = \eta_0/G_0$ values obtained for the four cell lines described above. The cellular data show viscosities ranging from 10 to 90 Pa s and elastic moduli between 30 and 80 Pa. Notably, the highly invasive epithelial cells (MDA-MB-231) exhibit lower viscosity compared to their healthy or less invasive counterparts [39]. For the three breast cell lines, MRS yields viscosity values about ten times higher than particle-tracking microrheology [26, 55, 56], and comparable to those of optical tweezers [57] (Supporting Information S4). The critical frequency $\omega_C$ observed for $L$ = 3.5 μm wires consistently remains on the order of 0.1 rad s$^{-1}$, while the relaxation times span an order of magnitude from 0.3 to 3 s. Among all the MRS data obtained by us on living cells, most of which remain unpublished, the results presented here are representative of the behavior of adherent cells commonly used in biophysics.

| Cell lines | Characteristics | $\omega_C$ (rad s$^{-1}$) | $\eta_0$ (Pa s) | $G_0$ (Pa) | $\tau$ (s) | $\omega_C\tau$ |
|---|---|---|---|---|---|---|
| MCF-10 | Human normal-like breast epithelial cells | 0.09 ± 0.04 | 46 ± 13 | 79 ± 7 | 0.6 ± 0.2 | 0.054 |
| MCF-7 | Low-invasive tumorigenic breast epithelial cells | 0.04 ± 0.03 | 87 ± 16 | 33 ± 6 | 2.7 ± 0.9 | 0.108 |
| MDA-MB-231 | High-invasive tumorigenic breast epithelial cells | 0.3 ± 0.08 | 12 ± 8 | 39 ± 6 | 0.3 ± 0.2 | 0.090 |
| HeLa | Human cervical cancer | 0.09 ± 0.07 | 28 ± 10 | 48 ± 5 | 0.6 ± 0.3 | 0.054 |

**Table I**: Median values and standard errors for critical frequency $\omega_C$, zero-shear shear viscosity $\eta_0$, elastic modulus $G_0$ and relaxation time $\tau$ determined from MRS of MCF-10A, MCF-7, MDA-MB-231 and HeLa cells. The values for $\omega_C$ and $\eta_0$ are derived for wires of length $L$ = 3.5 ± 1 μm *via* Eq. 1. The last column indicates that the product $\omega_C\tau$ remains below 1. For comparison, passive particle-tracking microrheology yielded average viscosity values of 3.55 ± 0.95 Pa s for MCF-10A, 0.51 ± 0.21 Pa s for MCF-7, and 0.83 ± 0.42 Pa s for MDA-MB-231 [11, 55, 56, 58]. Using optical tweezers, zero-shear viscosities of 18 Pa s and 13 Pa s were reported for MCF-10A and MDA-MB-231, respectively [57] (Supporting Information S4).

## III.2 – Velocity fields and gradients around a rotating wire

COMSOL Multiphysics three-dimensional simulations were performed using both a Newtonian fluid and a viscoelastic Oldroyd-B fluid, under conditions corresponding to steady-state wire rotation [36]. The Oldroyd-B model is suitable for describing viscoelastic fluids and reduces to the upper-convected Maxwell model when the solvent viscosity is set to zero, which was the simplifying assumption adopted here [59]. As a first step, we verified that in the linear flow regime, both models yielded identical velocity fields for the viscoelastic fluid, as shown in Supporting Information S5. The results demonstrate good agreement between the two simulations. It should also be noted that for the viscoelastic model, the mesh size of the simulation domain was increased compared with that used for the viscous fluid. Despite this adjustment, the computation time increased markedly, by a factor of ten between the two simulations. Given that the flow field remains nearly unchanged except in the immediate vicinity of the simulated wire, partly due to the mesh refinement, we chose to retain the Newtonian model, as it provides comparable results with a substantially lower computational cost.

Specifically, the wire was modeled as a cylinder void in the center of a cubic box, 20 times larger than the wire length. Fig. 2a displays 3D visualizations of iso-velocity surfaces of the steady-state flow field resulting from a rotating wire. The wire, measuring 3.5 µm in length and 0.70 µm in diameter, rotates about the $z$-axis at an angular frequency of 0.1 rad s⁻¹. The maximum velocity in this simulation is hence $v_{Max} = \omega L/2 = 0.175$ µm s⁻¹. The images from left to right illustrate the velocity fields for $v/v_{Max}$ = 30%, 50%, and 85%. Figs. 2b-d present the 2D projections of the previous 3D simulation data in the $(xOy)$, $(yOz)$, and $(xOz)$ planes, respectively. The 2D profiles exhibit pronounced anisotropies in velocity, and characteristic dumbbell-shaped patterns in two of the three panels. There, the maximum velocities occur in the wire tip. In the plane perpendicular to the wire, the velocity field is close to zero, as illustrated in Fig. 2c.

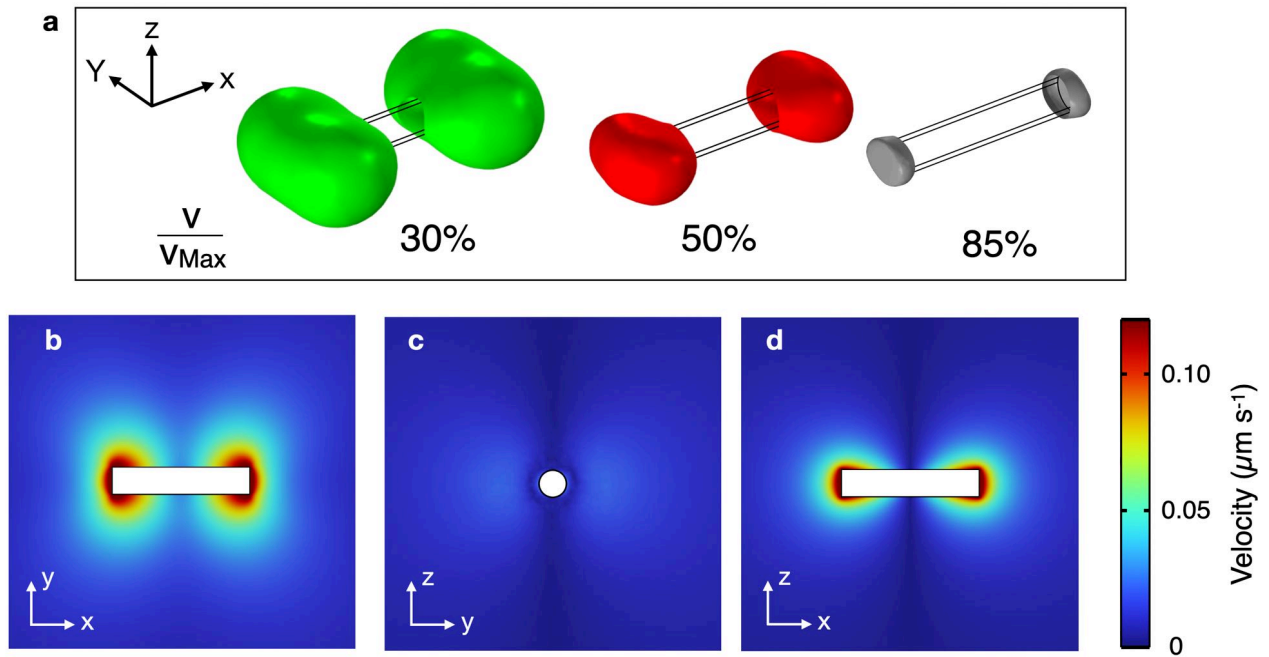

**Figure 2**: **a)** 3D-representations of iso-velocity surfaces of the flow field around a rotating wire ($L$ = 3.5 μm, $D$ = 0.70 μm,) obtained from COMSOL finite element modeling. The wire rotates at an angular frequency of 0.1 rad s⁻¹ around the $z$-axis, resulting in a maximum tip velocity $v_{Max}$ of 0.175 μm s⁻¹. The three images (from left to right) correspond to velocity ratios of $v/v_{Max}$ = 30%, 50%, and 85% respectively. **b,c,d)** 2D-representations of the velocity fields in the ($xOy$), ($yOz$), and ($xOz$) planes, respectively in the same conditions as in a). In the ($xOy$), and ($xOz$) planes, the velocity profiles display characteristic dumbbell-shaped patterns.

Figs. 3a-c illustrate the variation of the velocity moduli noted $v$ as functions of the distances $x$, $y$, and $z$ from the wire tip, respectively. In Supporting Information S6, we verify that the dominant components are aligned with the $y$-axis, and subsequently we identify the modulus of the velocities $v(x)$, $v(y)$, and $v(z)$ with their components along the $y$-axis, $v_y(x)$, $v_y(y)$, and $v_y(z)$. The simulations reveal a rapid velocity decrease, characterized by the initial velocity $v_{Max}$ and penetration lengths noted $\delta_{v,x}$, $\delta_{v,y}$ and $\delta_{v,z}$ for each of the three directions. The penetration lengths were calculated using two methods. In the first approach, they were determined as the intersection points of the initial linear region of the velocity profiles (i.e. near the body of the wire) with the distance axis. In the second approach, the profiles were fitted with a stretched exponential function, which proved to be a suitable model (Supporting Information S7). Both methods yielded similar results. Table II shows that these characteristic lengths increase progressively from $\delta_{v,z}$ = 0.18 μm in the $z$-direction to $\delta_{v,y}$ = 0.77 μm in the $y$-direction. In the present case, the $\delta_v$'s remain below 1 μm, that is smaller than the wire length and on the order of the wire diameter.

Figs. 3d-f depict the velocity gradients $\dot{\gamma}_x$, $\dot{\gamma}_y$ and $\dot{\gamma}_z$ obtained by differentiating the corresponding velocity fields. The velocity gradient exhibits a similar trend to the velocity profiles, rapidly decreasing with distance from the wire. A systematic analysis of the $\dot{\gamma}$-functions *versus* distance leads to the determination of the maximum shear rates $\dot{\gamma}_{Max,x}$, $\dot{\gamma}_{Max,y}$ and $\dot{\gamma}_{Max,z}$ and the penetration lengths $\delta_{\dot{\gamma},x}$, $\delta_{\dot{\gamma},y}$ and $\delta_{\dot{\gamma},z}$ along the three directions (Table II). The maximum shear rates are in the range 0.2 s$^{-1}$ (along $Oy$) to 0.8 s$^{-1}$ (along $Oz$). Once combined with intracellular relaxation times $\tau$ from Table I, these values indicate that, for the conditions tested here, the product $\dot{\gamma}\tau$ is less than one, hence in the linear regime of the rheological response. Furthermore, it can be noted that:

$$\dot{\gamma}_{Max,x} \approx v_{Max}/\delta_{v,x}; \; \dot{\gamma}_{Max,y} \approx v_{Max}/\delta_{v,y}; \; \dot{\gamma}_{Max,z} \approx v_{Max}/\delta_{v,z} \qquad (3)$$

where $v_{Max}$, $\delta_{v,x}$, $\delta_{v,y}$ and $\delta_{v,y}$ were retrieved from the velocity field analysis. The shear rate penetration lengths are here useful as they allow us to estimate the volume of cytoplasm, $V_{Shear}$, that is effectively sheared during a $2\pi$-rotation of the wire. This volume corresponds approximately to that of a flat cylinder with a diameter of $L + 2\delta_{\dot{\gamma},x}$ and a height of $D + 2\delta_{\dot{\gamma},z}$ leading to the expression: $V_{Shear} = \pi(L/2 + \delta_{\dot{\gamma},x})^2(D + 2\delta_{\dot{\gamma}z})$. For the simulation considered here and after subtracting the wire own volume, we obtain a sheared volume of $V_{Shear}$ = 16 fL, which represents approximately 1% of the total cell volume [60]. The relatively low penetration length along the $z$-axis ($\delta_{\dot{\gamma},z}$ = 0.28 μm) suggests negligible influence from cortical actin and membrane boundaries for most wires lying flat within the cytoplasm [61]. Taken together, these results underscore the localized nature of intracellular mechanical measurements obtained with MRS, and delineate the interaction range between the magnetic wire and its surrounding cytoplasm.

Velocity fields

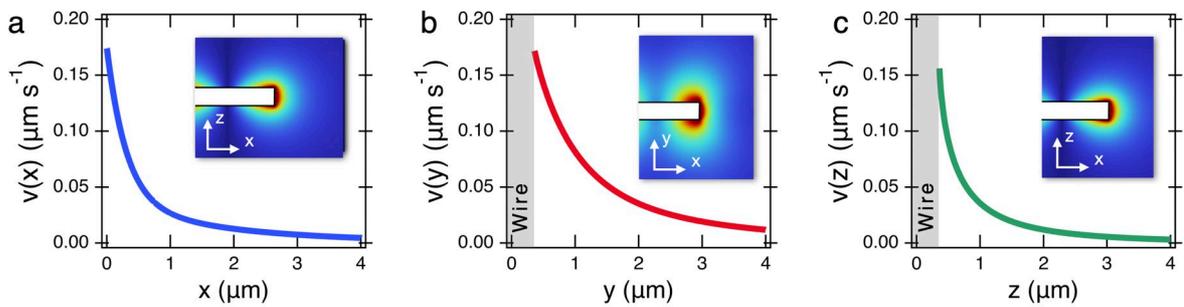

Velocity gradients

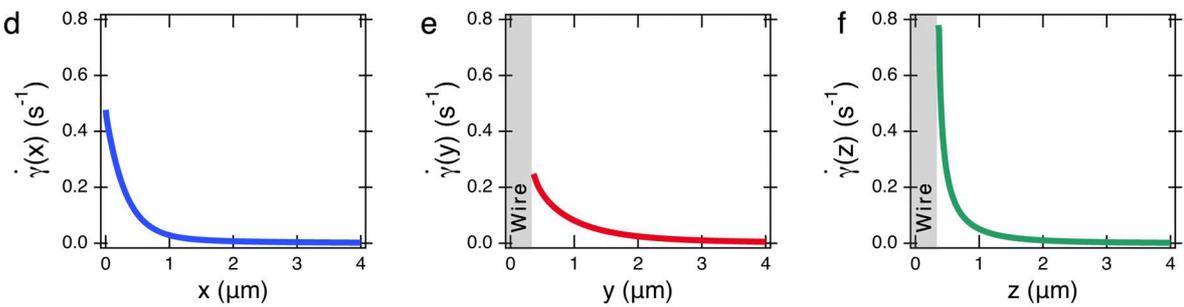

*Figure 3*: **a,b,c)** Velocity fields as a function of the distance from the wire along the $x$-, $y$-, and $z$-directions respectively. The velocities presented here correspond to the components along the $y$-axis (Supporting Information S6). Insets: images of the contours in the $(xOz)$, $(xOy)$ and $(xOz)$ planes. **d,b,c)** Velocity gradients $\dot{\gamma}_x$, $\dot{\gamma}_y$ and $\dot{\gamma}_z$ as a function of the distance calculated from the velocity fields in a-c), respectively. The grey areas in b), c) e) and f) represent the wire contour.

| Directions | $v_{Max}$ (μm s$^{-1}$) | $\delta_v$ (μm) | $\dot{\gamma}_{Max}$ (s$^{-1}$) | $\delta_{\dot{\gamma}}$ (μm) |
|---|---|---|---|---|
| $x$ | 0.175 | 0.44 ± 0.04 | 0.45 ± 0.03 | 0.42 ± 0.03 |
| $y$ | 0.175 | 0.78 ± 0.03 | 0.24 ± 0.02 | 0.48 ± 0.04 |

| $z$ | 0.175 | 0.20 ± 0.02 | 0.80 ± 0.10 | 0.28 ± 0.02 |
|---|---|---|---|---|

**Table II**: Values of the maximum velocity $v_{Max}$, maximum shear rate $\dot{\gamma}_{Max}$, and penetration lengths $\delta_v$ (resp. $\delta_{\dot{\gamma}}$) associated with the profiles shown in Figs. 3a-c (resp. Fig. 3d-f). Of note, the penetration lengths are smaller than the diameter of the wire, indicating a marked localization of the flow at the wire ends. The error bars reflect the variations due to the different adjustment approaches implemented for the velocity and gradient profiles.

## III.3 – Shear deformation during wire rotation

In this section, we examine the deformation experienced by a fluid element in the vicinity of the wire during its rotation. Due to symmetry, we limit our analysis to a half rotation of the wire, between the angle $\theta$ = -$\pi$/2 and $\pi$/2 (panels A and F in Fig.4a). The fluid elements considered here are chosen based on the previous outcomes, *i.e.* along the $x$- and $z$-axes where shear rates are highest. In Fig. 4a, the wire rotates clockwise at an angular frequency of $\omega$ = 0.1 rad s$^{-1}$, and the observation region is represented by the arrow. The panels labeled A to F correspond to wire orientations of $\theta$ = -$\pi$/2, -$\pi$/4, -$\pi$/18, $\pi$/18, $\pi$/4, and $\theta$ = $\pi$/2, respectively. Fig. 4b shows the evolution of the shear rate $\dot{\gamma}(\theta, x)$ as a function of angle and distance, with three representative cases: $x$ = 0, $\delta_{v,x}$ and $2\delta_{v,x}$. When the fluid volume comes close to the wire at the midpoint of its trajectory ($\theta$ = 0), the shear rate exhibits a broad maximum. This maximum can be schematized as a rectangular pulse function with a maximum value of 0.4 s$^{-1}$ and a duration of 3.5 s (shaded area). Outside the pulse region, the shear rate decreases rapidly, reaching values as low as 10$^{-3}$ s$^{-1}$ when the wire is perpendicular to the observation zone ($\theta$ = $\pm$ $\pi$/2). With increasing distance from the wire, the shear rate decreases significantly, leading to broadening of initial pulse function. Fig. 4c compiles this behavior up to distances of 2 µm showing a decreasing pattern consistent with the results in Fig. 3d. By integrating the $\dot{\gamma}(x, \theta)$ functions over the rotation angle for different $x$-values, we determine the total deformation $\gamma_T(x) = \int_{-\pi/2\omega_C}^{\pi/2\omega_C} \dot{\gamma}(x,t)dt$ experienced by a fluid element in the vicinity of the wire. The resulting total deformation value $\gamma_T(x)$ reaches approximately 2.4 close to the wire, decreasing sharply with distance to about $\gamma_T(x) = 0.1$ at $x = 2$ μm.

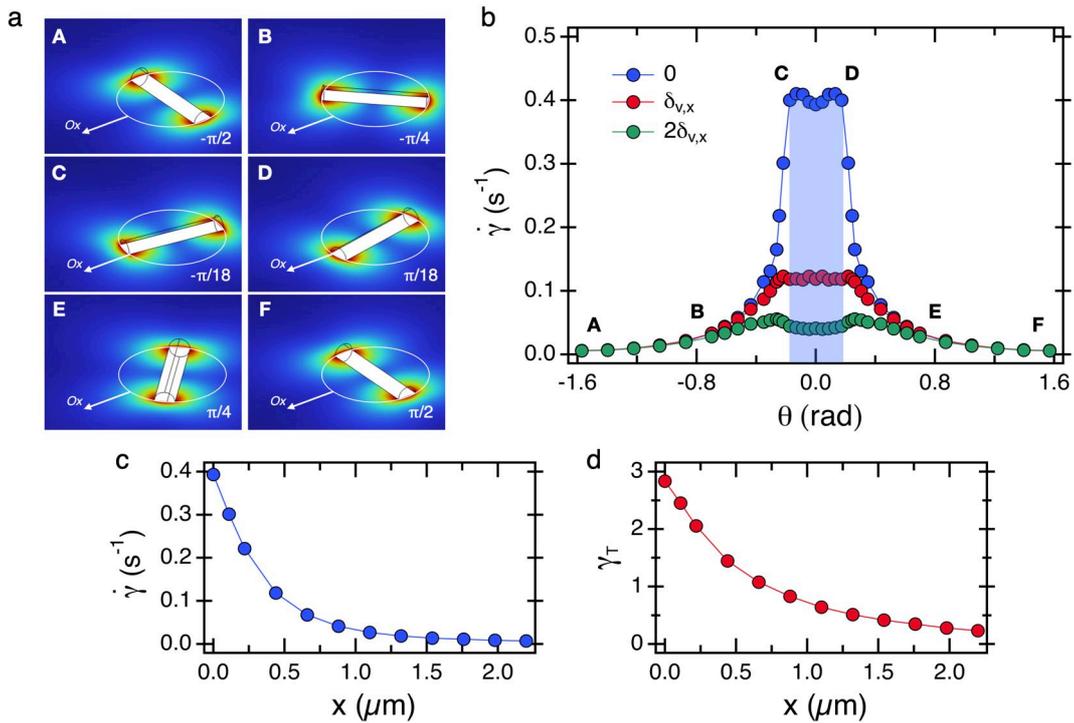

**Figure 4**: **a)** The configuration used in the current simulation focuses on a fluid element along the $x$-axis (white arrow) as the wire rotates clockwise from $\theta$ = -π/2 to π/2 at an angular frequency of $\omega_C$ = 0.1 rad s$^{-1}$. The panels A to F correspond to wire orientations of $\theta$ = -π/2, -π/4, -π/18, π/18, π/4, and $\theta$ = π/2, respectively. **b)** Evolution of the shear rate over the angle $\theta$ during a half rotation of the wire. The $\dot{\gamma}(x,\theta)$-functions are evaluated at various distances from the wire, ranging from 0 to 2.2 μm. The figure illustrates three representative cases: $x$ = 0, 1.1 μm, and 2.2 μm, corresponding to 0, 2.5 and 5 times the penetration length $\delta_{v,x}$. The shaded area represents an equivalent rectangular pulse function. **c)** Shear rate $\dot{\gamma}(x,\theta = 0)$ as a function of the distance as determined from the shear rate values at $\theta$ = 0. **d)** Total deformation $\gamma_T(x)$ experienced by the fluid element in the vicinity of the wire. The deformation is found to decrease rapidly with the distance, confirming the high localization of the sheared volume.

Fig. 5 displays the deformation data following the same analysis for a sample element along the $z$-axis, the wire undergoing a half-rotation. Fig. 5b presents the shear rate *versus* $\theta$ at various distances from the wire, namely at $D/2$ ($z = 0.35$ µm), $D/2 + \delta_{v,z}$ ($z = 0.55$ µm), $D/2 + 2\delta_{v,z}$ ($z = 0.75$ µm) and $D/2 + 4\delta_{v,z}$ ($z = 1.15$ µm). The previous rectangular pulse function is replaced by a pronounced peak observed when the wire reaches the midpoint of its trajectory. Figs. 5c and 5d show the variations in shear rate $\dot{\gamma}(z, \theta = 0)$ and total deformation $\gamma_T(z) = \int_{-\pi/2\omega_C}^{\pi/2\omega_C} \dot{\gamma}(z,t)dt$ as a function of distance from the wire. As before, the same trends are observed, with both quantities decreasing rapidly with distance.

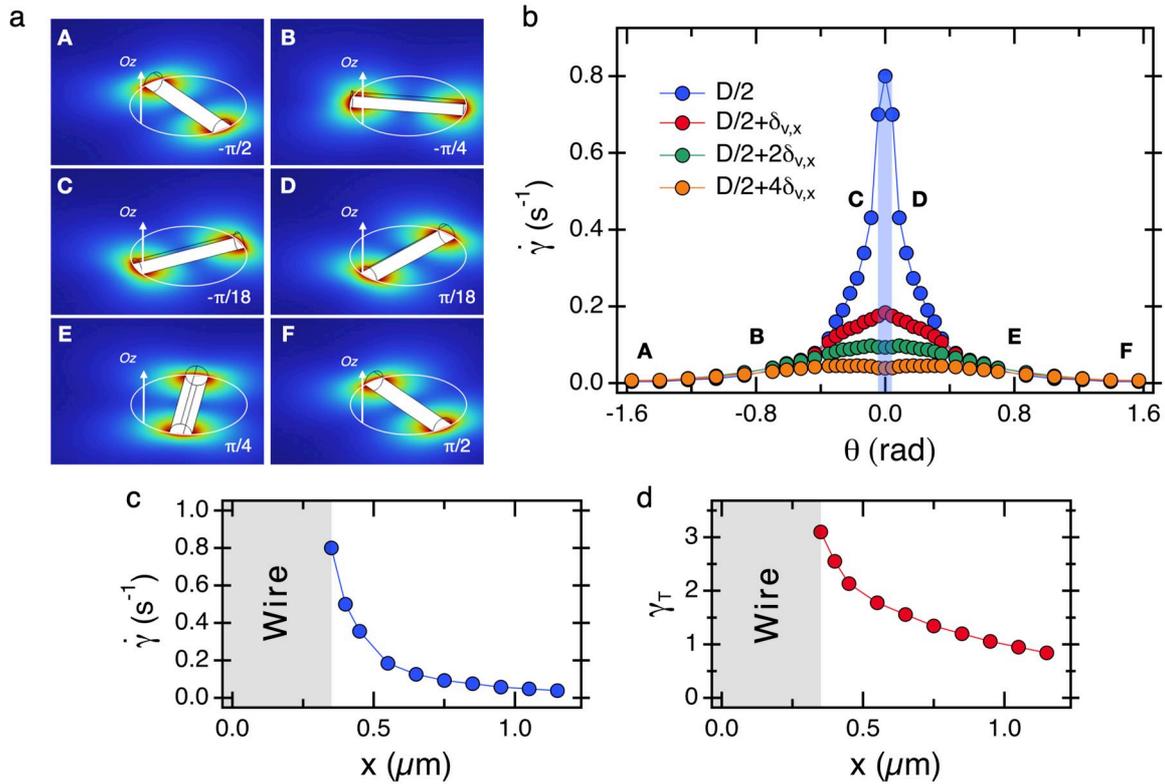

**Figure 5**: Same as in Fig. 4 for a fluid element along the $z$-axis. **a)** Schematical configuration of the wire orientation with respect to the tested volume. **b)** Shear rate $\dot{\gamma}(z, \theta)$ versus $\theta$ during a half rotation of the wire. The evaluations are made at $D/2$ ($z = 0.35$ µm), $D/2 + \delta_{v,z}$ ($z = 0.55$ µm), $D/2 + 2\delta_{v,z}$ ($z = 0.75$ µm) and $D/2 + 4\delta_{v,z}$ ($z = 1.15$ µm) from the wire. The shaded area represents an equivalent rectangular pulse function. **c)** Shear rate $\dot{\gamma}(z, \theta = 0)$ as a function of the distance as determined from the shear rate values at $\theta = 0$. **d)** Total deformation $\gamma_T(z)$ experienced by the fluid element in the vicinity of the wire during the half rotation of the wire.

In conclusion, during rotation, a fluid element near the wire end experiences a shear rate pulse function $\dot{\gamma}(t)$ at each half-period $\pi/\omega$. Outside this phase, both shear rates and deformations are significantly lower. The duration of this pulse varies depending on the direction and on the applied frequency. Based on the outcomes from Figs. 4 and 5, we found that the velocity gradient pulse duration goes as $\Delta t_x(\omega) \sim 1/3\omega$ in the $x$-direction and $\Delta t_z(\omega) \sim 1/8\omega$ in the $z$-direction. Furthermore, the fluid deformation is also confined, extending approximately 2 μm from the wire before becoming negligible. Close to the wire tip, however, the total deformation reaches relatively high values ($\gamma_T \sim$ 2-3), potentially indicating the occurrence of non-linear phenomena in this spatial region [34]. This non-linearity issue can also be assessed by evaluating the shear rates obtained from simulations under MRS cellular conditions.

## III.4 – Direct MRS-simulation comparison in the four investigated cell lines

In this section, we extend our investigation by simulating the response of a fluid of viscosity $\eta_0$ to wires of varying lengths. The results obtained in Section II.3 for an average length of 3.5 μm may possibly differ when considering the full range of experimental conditions tested. For these simulations, we used experimental data on wire length and diameter, on the critical frequency and on the viscosity collected across the 4 cell lines. First, we plotted the data $D(L)$, $L^*(L)$, $\omega_C(L^*)$ and $\eta_0(\omega_C)$ on double logarithmic scales and observed that all data sets followed a power-law relationship of the form $D(L) \sim L^\kappa$, $L^*(L) \sim L^\lambda$, $\omega_C(L^*) \sim L^{*\mu}$ and $\eta_0(\omega_C) \sim \omega_C{}^\nu$ where $\kappa$, $\lambda$, $\mu$, and $\nu$ are scaling exponents (Supporting Information S8). For each selected length, we evaluated $D$, $\omega_C$, and $\eta_0$ using the fitted scaling laws, and then used these values as inputs for computing velocity and shear rate fields through finite element simulations in COMSOL Multiphysics. This approach ensures that the simulations closely match experiments, allowing us to realistically reproduce the intracellular flow field.

Fig. 6 a-d illustrates the evolution of the penetration lengths $\delta_{v,x}(L)$, $\delta_{v,y}(L)$ and $\delta_{v,z}(L)$ *versus* wire length. In all cases, scaling laws with exponents 0.50 ± 0.10 are observed and are indicated by straight lines. Consistent with the data in Section II.3, the inequalities $\delta_{v,z} < \delta_{v,x} < \delta_{v,y}$ hold for all wire lengths, $\delta_{v,x}$ and $\delta_{v,z}$ remaining inferior to the wire diameter [61]. We then quantified the sheared volume fraction ($V_{Shear}/V_{Cell}$) as a function of wire length (Figs. 6e-h). Again, a scaling behavior is observed, with an exponent of 2.1 ± 0.1. For cell volumes, we used literature values: $V_{Cell}$(MCF-10A) = 2728 µm³, $V_{Cell}$(MCF-7) = 2397 µm³, and $V_{Cell}$(HeLa) = 2425 µm³ [62, 63]. For MDA-MB cells, an intermediate value between those of MCF-10A and MCF-7 was assumed ($V_{Cell}$ (MDA-MB) = 2500 µm³). The grayed-out segments represent the 3-4 µm interval, where the wires are most numerous. Here, the volume proportion is approximately 1% for all cell lines, further confirming the minimal impact of wire rotation on the cell.

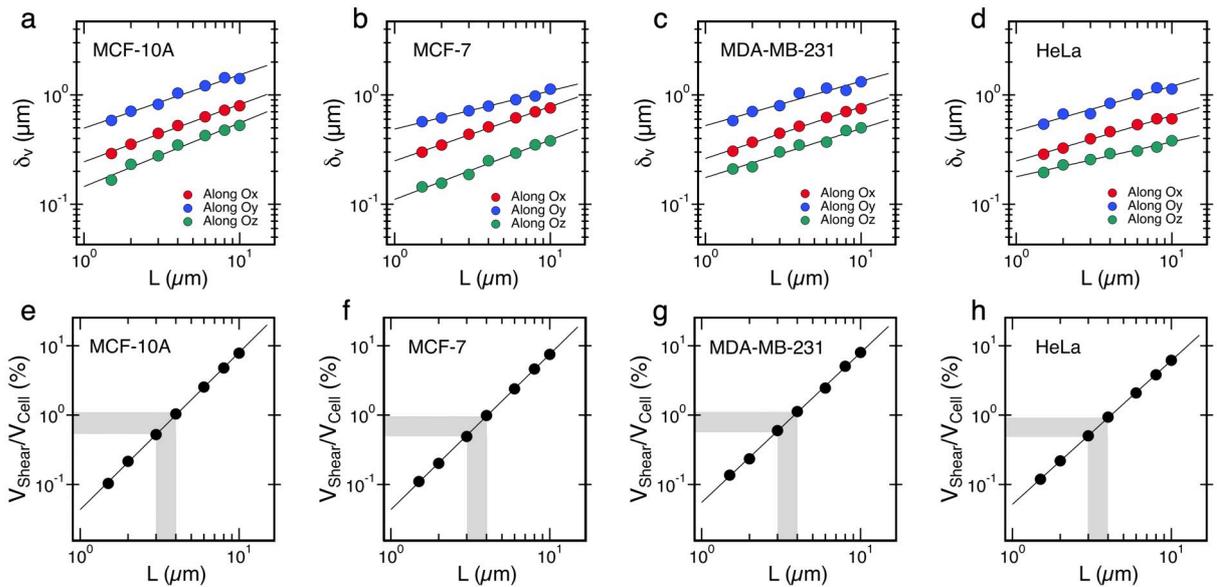

**Figure 6**: Penetration lengths $\delta_{v,x}(L)$, $\delta_{v,y}(L)$ and $\delta_{v,z}(L)$ as a function of the wire length $L$ for **a)** MCF-10A, **b)** MCF-7, **c)** MDA-MB-231 and **d)** HeLa cells obtained from finite elements simulations. The straight lines correspond to non-linear least squares fits using power laws, with exponents between 0.4 and 0.6. **e-f)** Sheared volumes calculated as $V_{Shear} = \pi(L/2 + \delta_{v,x})^2(D + 2\delta_{v,z})$ normalized to the average volume of the cells $V_{Cell}$, respectively. The $V_{Cell}$ values are taken from Refs. [62, 63]. The gray areas indicate that for wire lengths between 3 and 4 μm, the relative sheared volume remains around 1%.

Fig. 7a-d present the maximum reduced shear rate $\dot{\gamma}_{Max}\tau$ obtained for the four cell lines as a function of wire length in the $x$-, $y$-, and $z$-directions. As $\dot{\gamma}_{Max} \approx \omega_C L/2\delta_v$, the strong decreasing trend of $\dot{\gamma}_{Max}\tau$ with increasing $L$ is primarily governed by the dependence of the critical frequency and, to a lesser extent, by the increase in penetration length. As highlighted in introduction, the nonlinearity condition for the flow around the wire is given by $\dot{\gamma}_{Max}\tau > 1$. We find that this condition is met for wires shorter than $L = 1.8 \pm 0.2$ μm in MCF-10A, $L = 3.0 \pm 0.4$ μm in MCF-7, $L = 2.5 \pm 0.2$ μm in MDA-MB-231, and $L = 2.5 \pm 0.2$ μm in HeLa cells. Overall, the main conclusion is that for most wires studied in cells, the simulations indicate that the flow remains in the linear regime. In three out of four cell lines (MCF-10A, MDA-MB-231 and HeLa), only wires shorter than 2.5 μm may induce non-linear effects during rotation. Yet, these short wires represent a small (10-20%) fraction of the overall dataset (Supporting Information S3). For MCF-7 cells, caution is advised, as simulations predict non-linear effects for wire lengths below 3 μm. Here, we clearly demonstrate that comparing intracellular data with COMSOL simulations can enhance the accuracy of *in vitro* mechanical property assessments.

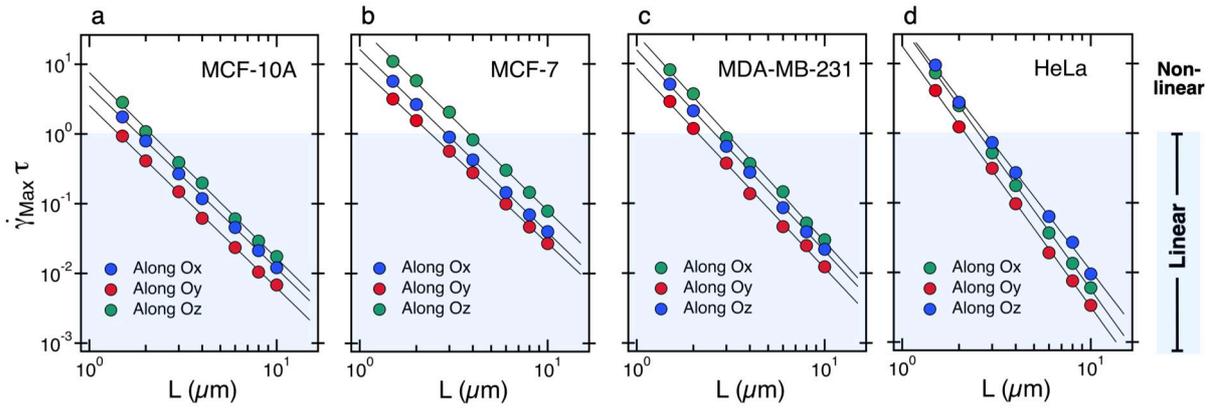

**Figure 7**: **a)** Product $\dot{\gamma}_{Max}\tau$ as a function of wire length $L$, obtained from COMSOL 3D simulations for MCF-10A epithelial cells. Here $\dot{\gamma}_{Max}$ denotes the maximum shear rate induced by wire rotation in the $x$-, $y$-, and $z$-directions (circles in blue, red and green respectively), and $\tau = \eta_0/G_0$ is the cytoplasm relaxation time. The straight lines represent nonlinear least-squares fits using power laws, with exponents between -2.6 and -2.7. The blue shaded area indicates the range corresponding to the linear regime of stationary flow. **b)** Same as in a) for MCF-7 epithelial cells. The scaling exponents are here comprised between -2.4 and -2.7. **c)** Same as in a) for MDA-MB-231 epithelial cells, with scaling exponents between -2.7 and -2.9. **d)** Same as in a) for HeLa cancer cells, with scaling exponents between -3.5 and -3.8.

## IV – Conclusion

COMSOL Multiphysics simulations were used to investigate the reliability of MRS in measuring cytoplasmic zero-shear viscosity. The simulations were conducted under standard experimental conditions typically used for MRS viscosity measurements—namely, a wire length of 3.5 μm and diameter of 0.7 μm—corresponding to the most frequently studied wires across the four tested cell lines: MCF-10A, MCF-7, MDA-MB-231, and HeLa. A crucial observation is that both shear velocity and shear gradient fields are highly localized around the wire under steady-state rotation. In the rotation plane, the velocity fields display characteristic dumbbell-shaped profiles. The penetration lengths, both parallel and perpendicular to the wire are smaller than the diameter, typically less than 1 μm. This implies that the sheared volume under standard conditions is approximately 20 femtolitres, or about 1% of the total cell volume. Furthermore, the shear velocity along the wire decreases linearly from the tips toward the center, meaning that the central region of the sheared volume is subjected to low shear stress. These outcomes explain why the mechanical stimulation produced by wire rotation, even over periods longer than one hour, does not significantly affect cellular physiology. At the characteristic rotation frequency $\omega_C$ used in experiments, each fluid element near the wire experiences a brief, rectangular pulse of shear. Importantly, the cytoplasmic relaxation time $\tau$ satisfies the condition $\omega_C \tau < 1$, indicating that the cytoplasm has sufficient time to relax between successive passes of the wire. A detailed analysis of the maximum shear rate, $\dot{\gamma}_{Max}$, near the wire further confirms that the product $\dot{\gamma}_{Max}\tau$ decreases sharply with wire length. For wire lengths greater than 3 μm, we observe that $\dot{\gamma}_{Max}\tau < 1$, ensuring linear flow regime in which the stress remains proportional to the shear rate. Under these conditions, the MRS viscosity corresponds to the zero-shear viscosity of the cytoplasm. These simulations also underscore the potential for nonlinear effects in the case of shorter wires, particularly those with lengths less than 2.5 μm. In conclusion, combining finite element simulations with MRS experimental data demonstrates that, as in complex fluids previously studied with this technique, the intrinsic rheological properties of the cytoplasm can be reliably determined. These insights reinforce the validity of MRS as a tool for probing intracellular viscosity, provided that appropriate experimental parameters are used.

## Supporting Information

Experimental setup: Magnetic rotating field device and associated equipment (S1) – Example of a field of view of a HeLa cells monolayer incubated with magnetic wires (S2) – Sampling wires according to their length (S3) – Review of the literature data on the intracellular viscosity obtained for MCF-10A, MCF-7 and MDA-MB-231 breast epithelial cells (S4) – Comparison between the Newton and Oldroyd-B models based on finite element simulations (S5) - Relative variation of the velocity component along $Oy$ and of the velocity modulus in the $x$-, $y$-, and $z$-directions (S6) – COMSOL Multiphysics finite element simulations of rotating wires in the conditions $L/D >> 1$ (S7) – Evidence of power-law relationships in MRS data across the studied cell lines for $D(L)$, $L^*(L)$, $\omega_C(L^*)$ and $\eta_0(\omega_C)$ (S3)

## Author contributions

Conceptualization: JFB – Methodology: CPM, FG, JFB – Software: CPM, FG – Validation: CPM, JFB – Formal analysis: JFB – Investigation: CPM, FG, JFB – Writing – original draft, review and editing: JFB – Writing – review and editing: JFB – Visualization: CPM, FG – Supervision: JFB – Funding Acquisition: JFB

## Conflict of interest

The authors have no competing interests to declare

## Ethics approval

Ethics approval is not required

## AI Disclosure Statement

Artificial intelligence was used to improve the clarity and English language of selected sentences and paragraphs. No AI-generated content was used for data analysis, scientific interpretation, or drafting original research findings. All scientific content was conceived, written, and validated by the authors.

## Data Accessibility Statement

Data supporting the findings of this study are provided in part in the Supplementary Information section. Movies (.avi) illustrating key steps in data acquisition, such as wire entry into living cells, wire rotation within the cytoplasm, and COMSOL simulations of wire rotation are available via the ZENODO repository and link: https://doi.org/10.5281/zenodo.17335446. COMSOL simulation files (.mph) for selected wire rotation cases, along with representative raw velocity field data (.txt) and cytoplasmic viscosity and elasticity data (.xlsx) for the four cell lines analyzed are also included and accessible through the same link.

## Acknowledgments

We thank Grégory Arkowitz, Robert Kuszelewicz, Benoit Ladoux, René-Marc Mège, Jean-Baptiste Manneville, Milad Radiom, Myriam Reffay and Gilles Tessier for fruitful discussions. We are also grateful to Marie Dessard for her contribution to the MRS experiments on breast epithelial cells. This research was supported in part by the Agence Nationale de la Recherche under the contracts ANR-21-CE19-0058-1 (MucOnChip) and ANR-24-CE42-6142-03 (ViscoMag2).

## TOC Image

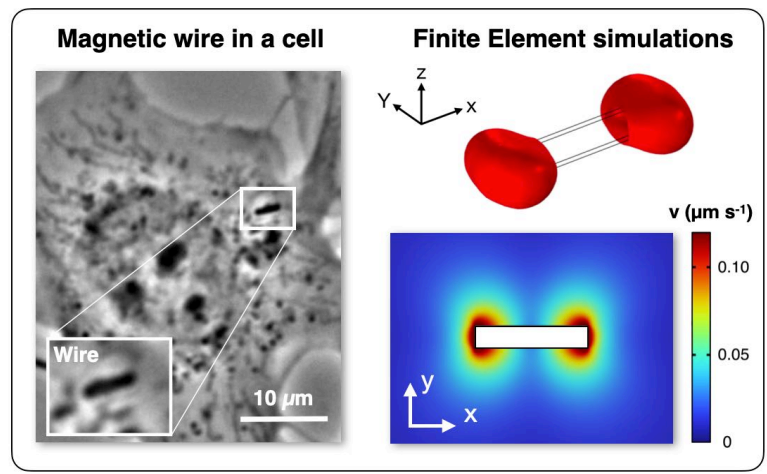

# References

[1] Alibert, C., Goud, B. & Manneville, J.-B. 2017 Are Cancer Cells Really Softer than Normal Cells? *Biol. Cell* **109**, 167-189. (doi:10.1111/boc.201600078).

[2] Lekka, M. 2016 Discrimination Between Normal and Cancerous Cells Using AFM. *BioNanoScience* **6**, 65-80.

[3] Liang, L., Song, X., Zhao, H. & Lim, C.T. 2024 Insights into the mechanobiology of cancer metastasis via microfluidic technologies. *APL Bioeng.* **8**, 021506. (doi:10.1063/5.0195389).

[4] Massey, A., Stewart, J., Smith, C., Parvini, C., McCormick, M., Do, K. & Cartagena-Rivera, A.X. 2024 Mechanical properties of human tumour tissues and their implications for cancer development. *Nature Rev. Phys.* **6**, 269–282. (doi:10.1038/s42254-024-00707-2).

[5] Guimarães, C.F., Gasperini, L., Marques, A.P. & Reis, R.L. 2020 The stiffness of living tissues and its implications for tissue engineering. *Nat. Rev. Mater.* **5**, 351-370. (doi:10.1038/s41578-019-0169-1).

[6] Suresh, S. 2007 Biomechanics and biophysics of cancer cells. *Acta Biomater.* **3**, 413-438. (doi:10.1016/j.actbio.2007.04.002).

[7] Wirtz, D., Konstantopoulos, K. & Searson, P.C. 2011 The physics of cancer: the role of physical interactions and mechanical forces in metastasis. *Nat. Rev. Cancer* **11**, 512-522. (doi:10.1038/nrc3080).

[8] Tanner, K. 2018 Perspective: The role of mechanobiology in the etiology of brain metastasis. *APL Bioeng.* **2**, 031801. (doi:10.1063/1.5024394).

[9] Markl, A., Nieder, D., Sandoval-Bojorquez, D., Taubenberger, A., Berret, J.-F., Yakimovich, A.s., Oliveros-Mata, E., Baraban, L. & Dubrovska, A. 2024 Heterogeneity of tumor biophysical properties and their potential role as prognostic markers. *Cancer Heterogen. Plasticity* **1**, 0011. (doi:10.47248/chp2401020011).

[10] Calzado-Martín, A., Encinar, M., Tamayo, J., Calleja, M. & San Paulo, A. 2016 Effect of Actin Organization on the Stiffness of Living Breast Cancer Cells Revealed by Peak-Force Modulation Atomic Force Microscopy. *ACS Nano* **10**, 3365-3374. (doi:10.1021/acsnano.5b07162).

[11] Li, Q.S., Lee, G.Y.H., Ong, C.N. & Lim, C.T. 2008 AFM Indentation Study of Breast Cancer Cells. *Biochem. Biophys. Res. Commun.* **374**, 609-613. (doi:10.1016/j.bbrc.2008.07.078).

[12] Nikkhah, M., Strobl, J.S., De Vita, R. & Agah, M. 2010 The Cytoskeletal Organization of Breast Carcinoma and Fibroblast Cells inside three Dimensional (3-D) Isotropic Silicon Microstructures. *Biomaterials* **31**, 4552-4561. (doi:10.1016/j.biomaterials.2010.02.034).

[13] Rother, J., Nöding, H., Mey, I. & Janshoff, A. 2014 Atomic Force Microscopy-Based Microrheology Reveals Significant Differences in the Viscoelastic Response between Malign and Benign Cell Lines. *Open Biol.* **4**, 140046. (doi:10.1098/rsob.140046).

[14] Stylianou, A., Lekka, M. & Stylianopoulos, T. 2018 AFM assessing of nanomechanical fingerprints for cancer early diagnosis and classification: from single cell to tissue level. *Nanoscale* **10**, 20930-20945. (doi:10.1039/C8NR06146G).

[15] Fischer, T., Hayn, A. & Mierke, C.T. 2020 Effect of Nuclear Stiffness on Cell Mechanics and Migration of Human Breast Cancer Cells. *Front. Cell Dev. Biol.* **8**, 393. (doi:10.3389/fcell.2020.00393).

[16] Blauth, E., Grosser, S., Sauer, F., Merkel, M., Kubitschke, H., Warmt, E., Morawetz, E.W., Friedrich, P., Wolf, B., Briest, S., et al. 2024 Different contractility modes control cell escape from multicellular spheroids and tumor explants. *APL Bioeng.* **8**, 026110. (doi:10.1063/5.0188186).

[17] Lekka, M., Laidler, P., Gil, D., Lekki, J., Stachura, Z. & Hrynkiewicz, A.Z. 1999 Elasticity of normal and cancerous human bladder cells studied by scanning force microscopy. *Eur. Biophys. J.* **28**, 312-316. (doi:10.1007/s002490050213).

[18] Gnanachandran, K., Kędracka-Krok, S., Pabijan, J. & Lekka, M. 2022 Discriminating bladder cancer cells through rheological mechanomarkers at cell and spheroid levels. *J. Biomech.* **144**, 111346. (doi:10.1016/j.jbiomech.2022.111346).

[19] Holuigue, H., Lorenc, E., Chighizola, M., Schulte, C., Varinelli, L., Deraco, M., Guaglio, M., Gariboldi, M. & Podestà, A. 2022 Force Sensing on Cells and Tissues by Atomic Force Microscopy. *Sensors* **22**, 2197.

[20] Yang, J., Antin, P., Berx, G., Blanpain, C., Brabletz, T., Bronner, M., Campbell, K., Cano, A., Casanova, J., Christofori, G., et al. 2020 Guidelines and definitions for research on epithelial-mesenchymal transition. *Nat Rev Mol Cell Biol* **21**, 341-352. (doi:10.1038/s41580-020-0237-9).

[21] Bastatas, L., Martinez-Marin, D., Matthews, J., Hashem, J., Lee, Y.J., Sennoune, S., Filleur, S., Martinez-Zaguilan, R. & Park, S. 2012 Afm Nano-Mechanics and Calcium Dynamics of Prostate Cancer Cells with Distinct Metastatic Potential. *Biochim. Biophys. Acta, Gen. Subj.* **1820**, 1111-1120. (doi:10.1016/j.bbagen.2012.02.006).

[22] Faria, E.C., Ma, N., Gazi, E., Gardner, P., Brown, M., Clarke, N.W. & Snook, R.D. 2008 Measurement of elastic properties of prostate cancer cells using AFM. *Analyst* **133**, 1498-1500. (doi:10.1039/b803355b).

[23] Lekka, M., Gil, D., Pogoda, K., Dulińska-Litewka, J., Jach, R., Gostek, J., Klymenko, O., Prauzner-Bechcicki, S., Stachura, Z., Wiltowska-Zuber, J., et al. 2012 Cancer cell detection in tissue sections using AFM. *Arch. Biochem. Biophys.* **518**, 151-156. (doi:10.1016/j.abb.2011.12.013).

[24] Tang, X., Zhang, Y., Mao, J., Wang, Y., Zhang, Z., Wang, Z. & Yang, H. 2022 Effects of substrate stiffness on the viscoelasticity and migration of prostate cancer cells examined by atomic force microscopy. *Beilstein J. Nanotech.* **13**, 560-569. (doi:10.3762/bjnano.13.47).

[25] Nguyen, A.V., Nyberg, K.D., Scott, M.B., Welsh, A.M., Nguyen, A.H., Wu, N., Hohlbauch, S.V., Geisse, N.A., Gibb, E.A., Robertson, A.G., et al. 2016 Stiffness of pancreatic cancer cells is associated with increased invasive potential. *Integr. Biol.* **8**, 1232-1245. (doi:10.1039/c6ib00135a).

[26] Li, Y., Schnekenburger, J. & Duits, M.H. 2009 Intracellular Particle Tracking as a Tool for Tumor Cell Characterization. *J. Biomed. Opt.* **14**, 064005. (doi:10.1117/1.3257253).

[27] Xu, W., Mezencev, R., Kim, B., Wang, L., McDonald, J. & Sulchek, T. 2012 Cell Stiffness Is a Biomarker of the Metastatic Potential of Ovarian Cancer Cells. *PLoS One* **7**, e46609. (doi:10.1371/journal.pone.0046609).

[28] Corbin, E.A., Kong, F., Lim, C.T., King, W.P. & Bashir, R. 2015 Biophysical properties of human breast cancer cells measured using silicon MEMS resonators and atomic force microscopy. *Lab. Chip* **15**, 839-847. (doi:10.1039/C4LC01179A).

[29] Dumitru, A.C., Mohammed, D., Maja, M., Yang, J., Verstraeten, S., Del Campo, A., Mingeot-Leclercq, M.P., Tyteca, D. & Alsteens, D. 2020 Label-Free Imaging of Cholesterol Assemblies Reveals Hidden Nanomechanics of Breast Cancer Cells. *Adv. Sci.* **7**, 2002643. (doi:10.1002/advs.202002643).

[30] Ketene, A.N., Schmelz, E.M., Roberts, P.C. & Agah, M. 2012 The Effects of Cancer Progression on the Viscoelasticity of Ovarian Cell Cytoskeleton Structures. *Nanomedicine* **8**, 93-102. (doi:10.1016/j.nano.2011.05.012).

[31] Nematbakhsh, Y., Pang, K.T. & Lim, C.T. 2017 Correlating the Viscoelasticity of Breast Cancer Cells with their Malignancy. *Converg. Sci. Phys. Oncol.* **3**, 034003. (doi:10.1088/2057-1739/aa7ffb).

[32] Rebelo, L.M., de Sousa, J.S., Mendes, J. & Radmacher, M. 2013 Comparison of the viscoelastic properties of cells from different kidney cancer phenotypes measured with atomic force microscopy. *Nanotechnology* **24**, 55102. (doi:10.1088/0957-4484/24/5/055102).

[33] Rianna, C. & Radmacher, M. 2017 Comparison of viscoelastic properties of cancer and normal thyroid cells on different stiffness substrates. *Eur. Biophys. J.* **46**, 309-324. (doi:10.1007/s00249-016-1168-4).

[34] Larson, R.G. 1998 *The Structure and Rheology of Complex Fluids*. New York, Oxford University Press.

[35] Squires, T.M. & Mason, T.G. 2010 Fluid Mechanics of Microrheology. *Annu. Rev. Fluid Mech.* **42**, 413-438. (doi:10.1146/annurev-fluid-121108-145608).

[36] COMSOL Multiphysics® v. 6.2. COMSOL AB 2019 Stockholm, Sweden

[37] Crick, F.H.C. & Hughes, A.F.W. 1950 The Physical Properties of Cytoplasm - A Study by Means of the Magnetic Particle Method .1. Experimental. *Exp. Cell Res.* **1**, 37-80. (doi:10.1016/0014-4827(50)90048-6).

[38] Berret, J.-F. 2016 Local Viscoelasticity of Living Cells Measured by Rotational Magnetic Spectroscopy. *Nat. Commun.* **7**, 10134. (doi:10.1038/ncomms10134).

[39] Dessard, M., Manneville, J.-B. & Berret, J.-F. 2024 Cytoplasmic viscosity is a potential biomarker for metastatic breast cancer cells. *Nanoscale Adv.* **6**, 1727-1738. (doi:10.1039/d4na00003j).

[40] Colby, R.H., Dealy, J.M., Morris, J., Morrison, F. & Vlassopoulos, D. 2013 Official symbols and nomenclature of The Society of Rheology. *J. Rheol.* **57**, 1047-1055. (doi:10.1122/1.4811184).

[41] Bonfanti, A., Kaplan, J.L., Charras, G. & Kabla, A. 2020 Fractional viscoelastic models for power-law materials. *Soft Matter* **16**, 6002-6020. (doi:10.1039/D0SM00354A).

[42] de Gennes, P.G. 1971 Reptation of a Polymer Chain in the Presence of Fixed Obstacles. *J. Chem. Phys.* **55**, 572-579. (doi:10.1063/1.1675789).

[43] Pusey, P.N. & van Megen, W. 1986 Phase behaviour of concentrated suspensions of nearly hard colloidal spheres. *Nature* **320**, 340-342. (doi:10.1038/320340a0).

[44] Chanteau, B., Fresnais, J. & Berret, J.-F. 2009 Electrosteric Enhanced Stability of Functional Sub-10 Nm Cerium And Iron Oxide Particles in Cell Culture Medium. *Langmuir* **25**, 9064-9070. (doi:10.1021/la900833v).

[45] Safi, M., Yan, M.H., Guedeau-Boudeville, M.A., Conjeaud, H., Garnier-Thibaud, V., Boggetto, N., Baeza-Squiban, A., Niedergang, F., Averbeck, D. & Berret, J.-F. 2011 Interactions between Magnetic Nanowires and Living Cells: Uptake, Toxicity, and Degradation. *ACS Nano* **5**, 5354-5364. (doi:10.1021/nn201121e).

[46] Schneider, C.A., Rasband, W.S. & Eliceiri, K.W. 2012 NIH Image to ImageJ: 25 Years of Image Analysis. *Nat. Methods* **9**, 671-675. (doi:10.1038/nmeth.2089).

[47] Chevry, L., Sampathkumar, N.K., Cebers, A. & Berret, J.F. 2013 Magnetic Wire-Based Sensors for the Microrheology of Complex Fluids. *Phys. Rev. E* **88**, 062306. (doi:10.1103/PhysRevE.88.062306).

[48] Radiom, M., Oikonomou, E.K., Grados, A., Receveur, M. & Berret, J.-F. 2022 Probing DNA-Amyloid Interaction and Gel Formation by Active Magnetic Wire Microrheology. In *Bacterial Amyloids: Methods and Protocols* (eds. V. Arluison, F. Wien & A. Marcoleta), pp. 285-303. New York, NY, Springer US.

[49] Tokarev, A., Luzinov, I., Owens, J.R. & Kornev, K.G. 2012 Magnetic Rotational Spectroscopy with Nanorods to Probe Time-Dependent Rheology of Microdroplets. *Langmuir* **28**, 10064-10071. (doi:10.1021/la3019474).

[50] Tokarev, A., Yatvin, J., Trotsenko, O., Locklin, J. & Minko, S. 2016 Nanostructured Soft Matter with Magnetic Nanoparticles. *Adv. Funct. Mater.* **26**, 3761-3782.

[51] Loosli, F., Najm, M., Chan, R., Oikonomou, E., Grados, A., Receveur, M. & Berret, J.-F. 2016 Wire-Active Microrheology to Differentiate Viscoelastic Liquids from Soft Solids. *ChemPhysChem* **17**, 4134-4143. (doi:10.1002/cphc.201601037).

[52] Frka-Petesic, B., Erglis, K., Berret, J.-F., Cebers, A., Dupuis, V., Fresnais, J., Sandre, O. & Perzynski, R. 2011 Dynamics of Paramagnetic Nanostructured Rods under Rotating Field. *J. Magn. Magn. Mater.* **323**, 1309-1313. (doi:10.1016/j.jmmm.2010.11.036).

[53] Helgesen, G., Pieranski, P. & Skjeltorp, A.T. 1990 Nonlinear Phenomena in Systems of Magnetic Holes. *Phys. Rev. Lett.* **64**, 1425-1428. (doi:10.1103/PhysRevLett.64.1425).

[54] Banks, H.T., Hu, S. & Kenz, Z.R. 2011 A Brief Review of Elasticity and Viscoelasticity for Solids. *Adv. Appl. Math. Mech.* **3**, 1-51. (doi:10.4208/aamm.10-m1030).

[55] Gal, N. & Weihs, D. 2012 Intracellular Mechanics and Activity of Breast Cancer Cells Correlate with Metastatic Potential. *Cell. Biochem. Biophys.* **63**, 199-209. (doi:10.1007/s12013-012-9356-z).

[56] Guo, M., Ehrlicher, A.J., Jensen, M.H., Renz, M., Moore, J.R., Goldman, R.D., Lippincott-Schwartz, J., Mackintosh, F.C. & Weitz, D.A. 2014 Probing the Stochastic, Motor-Driven Properties of the Cytoplasm Using Force Spectrum Microscopy. *Cell* **158**, 822-832. (doi:10.1016/j.cell.2014.06.051).

[57] Mandal, K., Asnacios, A., Goud, B. & Manneville, J.-B. 2016 Mapping Intracellular Mechanics on Micropatterned Substrates. *Proc. Natl. Acad. Sci.* **113**, E7159-E7168. (doi:10.1073/pnas.1605112113).

[58] Smelser, A.M., Macosko, J.C., O'Dell, A.P., Smyre, S., Bonin, K. & Holzwarth, G. 2015 Mechanical Properties of Normal versus Cancerous Breast Cells. *Biomech. Model. Mechanobiol.* **14**, 1335-1347. (doi:10.1007/s10237-015-0677-x).

[59] Larson, R.G. 1988 *Constitutive Equations for Polymer Melts and Solutions*, Butterworth-Heinemann.

[60] Rollin, R., Joanny, J.-F. & Sens, P. 2023 Physical basis of the cell size scaling laws. *eLife* **12**, e82490. (doi:10.7554/eLife.82490).

[61] Padding, J.T. & Briels, W.J. 2010 Translational and rotational friction on a colloidal rod near a wall. *J. Chem. Phys.* **132**, 054511. (doi:10.1063/1.3308649).

[62] McLaughlin, A.J., Kaniski, A.J., Matti, D.I. & Xhabija, B. 2023 Comparative Morphological Analysis of MCF10A and MCF7 Cells Using Holographic Time-lapse Microscopy. *Anticancer Res.* **43**, 3891-3896. (doi:10.21873/anticanres.16576).

[63] Zhao, L., Kroenke, C.D., Song, J., Piwnica-Worms, D., Ackerman, J.J. & Neil, J.J. 2008 Intracellular water-specific MR of microbead-adherent cells: the HeLa cell intracellular water exchange lifetime. *NMR Biomed.* **21**, 159-164. (doi:10.1002/nbm.1173).